\documentclass[12pt]{article}
\pdfoutput=1

\usepackage[utf8]{inputenc}
\usepackage[margin=1in]{geometry}
\usepackage[tbtags]{amsmath} 				
\usepackage{amssymb, mathrsfs}			
\usepackage{physics}
\usepackage{graphicx}			
\usepackage{tensor}				
\usepackage[makeroom]{cancel}	
\usepackage{bm}								
\usepackage[
      colorlinks=true,
      linkcolor=blue,
      urlcolor=blue,
      filecolor=black,
      citecolor=red,
      pdfstartview=FitV,
      pdftitle={},
      pdfauthor={Kevin Chen, Michael Gutperle},
      bookmarksopen=true
      ]{hyperref}

\def\Tr{{\rm Tr}}

\usepackage{sectsty}
\sectionfont{\large}

\renewcommand{\thefootnote}{\fnsymbol{footnote}}
\renewcommand{\thanks}[1]{\footnote{#1}}
\newcommand{\starttext}{
\setcounter{footnote}{0}
\renewcommand{\thefootnote}{\arabic{footnote}}}
\renewcommand\({\begin{equation}}		
\renewcommand\){\end{equation}}
\renewcommand{\epsilon}{\varepsilon}	

\numberwithin{equation}{section} 		

\numberwithin{equation}{section}

\long\def\symbolfootnote[#1]#2{\begingroup%
\def\thefootnote{\fnsymbol{footnote}}\footnote[#1]{#2}\endgroup}

\begin{document}
\setlength{\baselineskip}{16pt}

\starttext
\setcounter{footnote}{0}

\begin{flushright}
\today
\end{flushright}

\bigskip

\begin{center}

{\Large \bf  Holographic line defects in $F(4)$ gauged supergravity}

\vskip 0.4in

{\large  Kevin Chen and  Michael Gutperle}

\vskip 0.2in

{ \sl Mani L. Bhaumik Institute for Theoretical Physics} \\
{\sl Department of Physics and Astronomy }\\
{\sl University of California, Los Angeles, CA 90095, USA} 

\bigskip

\end{center}
 
\begin{abstract}
\setlength{\baselineskip}{16pt}
In this note we construct a  solution of six-dimensional  $F(4)$ gauged supergravity using   $AdS_2\times S^3$ warped over an interval as an ansatz. The solution is completely regular, preserves eight of the sixteen supersymmetries of the $AdS_6$ vacuum and  is a holographic realization of a line defect in a  dual  five-dimensional theory. We calculate the expectation value of the defect and the one-point function of the stress tensor in the presence of the defect using holographic renormalization.
\end{abstract}

\setcounter{equation}{0}
\setcounter{footnote}{0}

\newpage

\section{Introduction}

The study of five-dimensional superconformal theories (SCFTs) has been a very active field of research in recent years.  The fact that the theories are nonrenormalizable  implies that the definition of the theories in the UV is not straightforward. Examples of such theories were 
first constructed  utilizing decoupling limits of string theory \cite{Seiberg:1996bd,Morrison:1996xf,Intriligator:1997pq,Aharony:1997ju,Aharony:1997bh}. Holography is an important tool in studying CFTs and $AdS_6$ solutions dual to five-dimensional SCFTs have 
been found in massive IIA \cite{Brandhuber:1999np,Bergman:2012kr,Passias:2012vp} and T-duals in  type IIB \cite{Lozano:2012au,Lozano:2013oma,Kelekci:2014ima}. Recently, a large class of type IIB solutions were constructed in \cite{DHoker:2016ujz,DHoker:2016ysh,DHoker:2017mds,DHoker:2017zwj}\footnote{For earlier work on $AdS_6$ type IIB solutions see \cite{Apruzzi:2014qva,Kim:2015hya}.} which are 
dual to $d=5$ SCFTs related in the IR to  long quiver theories derived from $(p,q)$ five-brane webs \cite{Aharony:1997bh}. Apart from local operators, extended defect 
operators such as Wilson or t'Hooft lines, surface operators, and Janus-like defects are important objects which can be studied in QFT.  Of 
particular interest in SCFTs are defects which preserve a superconformal subalgebra of the original superconformal algebra.  Five-dimensional SCFTs have a unique superconformal algebra $F(4)$  \cite{Nahm:1977tg,Kac:1977em} and its subalgebras were classified in  \cite{VanderJeugt:1987ezi,Frappat}.  This analysis shows that superconformal defects should exist, such as a half-BPS Janus solution found in \cite{Gutperle:2017nwo}. 

In this paper we are constructing supergravity solutions corresponding to line defects preserving eight of the sixteen supersymmetries of $F(4)$, falling into a $D(2,1;2)\times SU(2)$ subsuperalgebra. One approach to construct holographic line defects is to consider probe branes \cite{Assel:2012nf}. Our aim is to construct  nonsingular supergravity solutions which correspond to the fully back-reacted solution which should describe the system  when the number of probe branes becomes large (see e.g.~\cite{DHoker:2007mci} for the case of Wilson line defects in $AdS_5\times S^5$). The fact that the ten-dimensional IIA and IIB  undeformed $AdS_6$ vacuum solutions are  already warped products 
 makes the construction of holographic defect solutions in ten dimensions quite challenging. In this paper we consider a simpler system,  namely   Romans' six-dimensional $F(4)$  gauged supergravity \cite{Romans:1985tw}. Recent results  \cite{Malek:2018zcz,Hong:2018amk,Malek:2019ucd} show  that any solution of this six-dimensional theory can be uplifted and embedded in the general IIB solutions  of  \cite{DHoker:2016ujz,DHoker:2016ysh,DHoker:2017mds}. This implies  that the solutions  in this paper lift to ten-dimensional holographic defect solutions. Recently, various supersymmetric solutions of  $F(4)$ supergravity without additional matter multiplets have been constructed in 
\cite{Kim:2019fsg,Suh:2018tul,Alday:2015jsa}. Examples of solutions of $F(4)$ gauged supergravity with matter couplings can be found in \cite{Gutperle:2017nwo,Suh:2018szn,Gutperle:2018axv,Hosseini:2018usu}.

The structure of the paper is as follows: in section \ref{sec2} we present the necessary background on  $F(4)$ gauged supergravity. In section \ref{sec3} we derive the nonsingular line defect solution using the BPS equations first derived in \cite{Dibitetto:2018gtk}. In section \ref{sec4} we perform some holographic calculations using the solution presented in section \ref{sec3}. In particular,  we calculate the on-shell action and the one-point function of the stress tensor, using holographic renormalization. Some implications of our solution and directions for future research are given in section \ref{sec5}. In the appendices we present our conventions and details of the calculation of the counterterms using the method of holographic renormalization.

\newpage

\section{$F(4)$ gauged supergravity}
\label{sec2}

In this section we review the features of $F(4)$ gauged supergravity \cite{Romans:1985tw} which will be relevant in this paper.
Six-dimensional $F(4)$ gauged supergravity contains the following bosonic fields:  a metric $G_{\mu\nu}$, a real scalar $\phi$, a 2-form gauge potential $B$, a non-Abelian $SU(2)$-valued vector field $A^i$ for $i=1,2,3$, and an Abelian vector field $A^0$.
The bosonic Lagrangian of the theory takes the following form,\footnote{See appendix \ref{appa} for our conventions regarding differential forms.}
\begin{align} 
\mathcal{L} =\, & R *_6 1 - 4 X^{-2} *_6 \dd{X} \wedge \dd{X}  - V(X) *_6 1 \nonumber \\
& - \frac{1}{2} X^4 *_6 H \wedge H - \frac{1}{2} X^{-2} \qty( *_6 F^i \wedge F^i + *_6 F \wedge F ) \nonumber \\
& - B \wedge \qty( \frac{1}{2} \dd{A^0} \wedge\dd{A^0}  + \frac{1}{\sqrt{2}} m B \wedge \dd{A^0} + \frac{1}{3} m^2 B \wedge B + \frac{1}{2} F^i \wedge F^i )  \label{eq-1-lagrangian}
\end{align}
where the field strengths derived from the potentials are given by
\begin{align}
H &= \dd{ B}  \nonumber \\
F^i &= \dd{A^i} + \frac{g}{2} \epsilon_{ijk} A^j \wedge A^k \nonumber \\
F &= \dd{A^0} + \sqrt{2} m B
\end{align}
and, for convenience, the scalar field $\phi$ has been redefined in terms of $X$ by
\( X = \exp( - \frac{1}{2 \sqrt{2}} \phi) \label{eq-1-Xdef} \)
Then the potential produced by the gauging of the supergravity is given by
\( V(X) = m^2 X^{-6} - 4 \sqrt{2} g m X^{-2} - 2 g^2 X^2 \)
which can be rewritten in terms of a superpotential $f(X)$ as
\begin{align}
V(X) &= 16 X^2 \qty(\partial_X f(X))^2 - 80 f(X)^2 \nonumber \\
f(X) &= \frac{1}{8} (m X^{-3} + \sqrt{2} g X)
\end{align}
The equations of motion following from the variation of the Lagrangian (\ref{eq-1-lagrangian}) are
\begin{align}
R_{\mu\nu} =\, & 4 X^{-2} \partial_\mu X \partial_\nu X + \frac{1}{4} V(X) G_{\mu\nu} + \frac{1}{4} X^4 \qty( \tensor{H}{_\mu^{\alpha\beta}} H^{\phantom \mu}_{\nu\alpha\beta} - \frac{1}{6} H^{\alpha\beta\gamma} H_{\alpha\beta\gamma} G_{\mu\nu} )  \nonumber \\
& + \frac{1}{2} X^{-2} \qty( \tensor{F}{_\mu^\alpha} F^{\phantom \mu}_{\nu\alpha} - \frac{1}{8} F^{\alpha\beta} F_{\alpha\beta} G_{\mu\nu}) + \frac{1}{2} X^{-2} \qty( \tensor{F}{^i_\mu^\alpha} \tensor{F}{^i_{\nu\alpha}} - \frac{1}{8} F^{i\alpha\beta} \tensor{F}{^i_{\alpha\beta}} G_{\mu\nu} ) \nonumber \\
\dd{ \qty(X^4 *_6 H ) } =\, & - \frac{1}{2} F \wedge F - \frac{1}{2} F^i \wedge F^i - \sqrt{2} m X^{-2} *_6 F \nonumber \\
\dd{ \qty(X^{-2} *_6 F ) } =\, & - F \wedge H \nonumber \\
\text{D} \qty(X^{-2} *_6 F^i) =\, &  - F^i \wedge H \nonumber \\
\dd{ \qty(X^{-1} *_6 \dd{X})} =\, & \frac{1}{8} X^{-2} \qty( *_6 F \wedge F + *_6 F^i \wedge F^i ) - \frac{1}{4} X^4 *_6 H \wedge H - \frac{1}{8} X \partial_X V(X) *_6 1 \label{eq-1-eom}
\end{align}
where $\text{D}$ is the gauge covariant derivative,
\( \text{D} F^i = \dd{ F^i} + g \epsilon_{ijk} A^j \wedge F^k \)
The supersymmetry variations of the fermionic fields can be expressed in terms of an $SU(2)$-doublet of symplectic-Majorana-Weyl Killing spinors $\zeta^a$ for $a=1, 2$ as
\begin{align}
\delta \psi^a_\mu =\, & \nabla_\mu \zeta^a + g A_\mu^i \tensor{(T^i)}{^a_b} \zeta^b - i f(X) \Gamma_\mu \Gamma_* \zeta^a  + \frac{X^2}{48}  H_{\nu\rho\sigma} \Gamma^{\nu\rho\sigma}  \Gamma_\mu \Gamma_* \zeta^a \nonumber \\
& + i \frac{X^{-1}}{16\sqrt{2}} \qty(\tensor{\Gamma}{_\mu^{\nu\rho}} - 6 \tensor{e}{_\mu^\nu} \Gamma^\rho ) \qty( F_{\nu\rho} \tensor{\delta}{^a_b} - 2 \Gamma_* F^i_{\nu\rho} \tensor{(T^i)}{^a_b} ) \zeta^b\\
\delta \chi^a =\, & X^{-1} \Gamma^\mu \partial_\mu X \zeta^a + 2i X \partial_X f(X) \Gamma_* \zeta^a - \frac{X^2}{24} H_{\mu\nu\rho} \Gamma^{\mu\nu\rho} \Gamma_* \zeta^a \nonumber \\
& -i \frac{X^{-1}}{8 \sqrt{2}} \Gamma^{\mu\nu} \qty( F_{\mu\nu} \tensor{\delta}{^a_b} - 2 \Gamma_* F_{\mu\nu}^i \tensor{(T^i)}{^a_b}) \zeta^b
\end{align}
where $\Gamma^m$ for $m=1, 2, \dotsc, 6$ generate the $(5+1)$-dimensional  Clifford algebra in an orthonormal frame and $\Gamma_* = \Gamma^{123456}$.
The $T^i = -i \sigma^i/2$ are the generators of $SU(2)$ satisfying $[T^i, T^j] = \epsilon_{ijk} T^k$. 

The space of inequivalent theories are labeled by the couplings $m$ and $g$, modulo the parameter rescaling $g \to a^{-1} g$,  $m \to a^3 m$ accompanied by appropriate field redefinitions.
The choice
\( g = \frac{3 m}{\sqrt{2}} \)
is a canonical choice, so that in  the supersymmetric $AdS_6$ vacuum the scalar takes the value $X = 1$.
We will make this choice throughout this paper, using $m$ in lieu of $g$.
The potential then takes the form
\( V(X) = m^2 \qty( X^{-6} - 12 X^{-2} -9 X^2) \) 

In   \cite{Cvetic:1999un} it was shown that six-dimensional $F(4)$ gauged supergravity  is a consistent nonlinear Kaluza-Klein reduction  of  the warped $AdS_6$ solutions of  type IIA massive supergravity. Recently an analogous statement  has been shown  \cite{Malek:2018zcz,Hong:2018amk}  for the warped $AdS_6\times S_2$ solutions of type IIB supergravity found in  \cite{DHoker:2016ujz,DHoker:2016ysh,DHoker:2017mds}. The fact that such a consistent truncation exists implies that any solution of  $F(4)$ gauged supergravity can be lifted to  ten-dimensional solutions, which have precise holographic duals. For example, the massive type IIA solution is dual to a $d=5$, $USp(N)$ gauge theory  for large $N$ \cite{Brandhuber:1999np}. Consequently, the defect solution we construct in section \ref{sec3}  also exists in the $AdS_6$ solutions in type massive  IIA and type IIB and corresponds to a line defect in the dual CFT.

\section{Defect solution}
\label{sec3}

In this section we find a nonsingular line defect solution by solving the BPS equations. An appropriate ansatz can be obtained by considering the unbroken subsuperalgebra of the superconformal algebra $F(4)$ suitable for a conformal line defect, namely $D(2,1;2)\times SU(2)$,  which has a bosonic part   $SO(2,1)\times SU(2)^3$. We can associate the $SO(2,1)$ with the global isometry of an $AdS_2$ factor. The three $SU(2)$ factors can be interpreted as the isometry $SO(4)\sim SU(2)\times SU(2)$ of a three sphere $S^3$ and unbroken $SU(2)$ R-symmetry. 
Consequently, the isometries are realized by  an   $AdS_2 \times S^3$ geometry warped over an interval $I_\alpha$,
\begin{align}\label{ansatza}
\dd{s}^2 &= e^{2U(\alpha)} \dd{s^2_{AdS_2}} + e^{2V(\alpha)} \dd{\alpha^2} + e^{2W(\alpha)} \dd{s^2_{S^3}} 
\end{align}
where $\dd{s^2_{AdS_2}}$ and $\dd{s^2_{S^3}}$ are unit-radius metrics.
Note that the warp factor $V$ is non-dynamical, but it is introduced because its gauge-fixing will turn out to simplify the BPS equations drastically.
The isometries and unbroken R-symmetry imply that all gauge fields have to vanish, but there can be a non-vanishing $B$ potential along the $AdS_2$ factor  and a nontrivial scalar profile,
\begin{align}
B &= b(\alpha)  \, \text{vol}_{AdS_2} \, , &
X &= X(\alpha) \, , & A^0 &= A^i = 0\, ,  \label{ansatzb} 
\end{align}
where  $\text{vol}_{AdS_2}$ is a unit-radius volume 2-form.

\subsection{BPS equations}
\label{sec-2-1}

The BPS equations for the ansatz (\ref{ansatza}) and (\ref{ansatzb}) have been derived  in \cite{Dibitetto:2018gtk}, where it was shown that solutions which preserve eight of the sixteen supercharges satisfy the following  system of first-order ordinary differential equations (ODEs),\footnote{We have set $L = 1$ in the equations of  \cite{Dibitetto:2018gtk}.}
\begin{align}
\theta' &= - e^V \sin(2\theta) (f - X \partial_X f) \nonumber \\
X' &= - \frac{1}{4} e^V X \cos(2\theta)^{-1} \qty( e^{-U} \sin(2\theta) + 2 \sin(2\theta)^2 f + (7 + \cos(4\theta)) X \partial_X f )  \nonumber \\
U' &= \frac{1}{4} e^V \cos(2\theta)^{-1} \qty( e^{-U} \sin (2 \theta) + (5 + 3 \cos(4\theta)) f + 6 \sin(2 \theta)^2 X \partial_X f ) \nonumber \\
W' &= - \frac{1}{4} e^V \cos(2\theta)^{-1} \qty( - e^{-U} \sin(2\theta)  + (-9 + \cos(4\theta)) f + 2 \sin(2\theta)^2 X \partial_X f ) \nonumber \\
b' &= - \frac{e^{V + 2U}}{X^2} \cos(2\theta)^{-1} \qty( e^{-U} + 2 \sin(2\theta) (f + 3 X \partial_X f) ) \nonumber \\
Y' &= \frac{Y}{8} e^V \cos(2\theta)^{-1} \qty( e^{-U} \sin(2\theta) + (5 + 3 \cos(4\theta)) f + 6 \sin(2\theta)^2 X \partial_X f  ) \label{eq-2-bps} 
\end{align}
where $'$ denotes the derivative with respect to $\alpha$. 
$Y$ and $\theta$ are functions related to the spinor parameters $\zeta^a$. 

The first three equations for $\theta'$, $X'$, and $U'$ should be treated as a coupled system of ODEs.
Once these are solved, the last three equations for $W'$, $b'$, and $Y'$ should be treated as three independent ODEs, the right-hand sides acting as inhomogenous terms.
In fact, assuming we have a solution of the first three equations, the solution for $W(\alpha)$ and $b(\alpha)$ is
\begin{align}
b &= b_0  - \frac{e^{2U}}{m} X \qty( e^{-U} + 2 \sin(2\theta)(f - X \partial_X f)) \nonumber \\
e^{-W} &= m r  \qty( e^{-U} \cos(2\theta)^{-1} + 2\tan(2\theta) (3f + X\partial_X f)) \label{eq-2-Wb}
\end{align}
where $b_0$ and $r$ are (real) integration constants.
$b_0$ is set to zero in order to satisfy the equations of motion.
$r$ can be interpreted as the $S^3$ radius.
The solution for $Y(\alpha)$ is inconsequential for our considerations in this paper, but for completion is
\( Y = Y_0 e^{U/2} \)
where $Y_0$ is a constant. 

To simplify the first three equations in \eqref{eq-2-bps}, we pick a gauge on the warp factor $V$ \cite{Dibitetto:2018gtk},
\( e^{-V} = \sin(2\theta) (f - X\partial_X f) \label{eq-2-V} \)
so that the first equation in \eqref{eq-2-bps} becomes $\theta' = -1$.
The associated integration constant involves constant shifts of $\alpha$, which has no physical consequence.
So we can set 
\( \theta(\alpha) = -\alpha \)
Then the two remaining equations become
\begin{align}
X' &= \frac{X}{4m \sin(2\alpha) \cos(2\alpha)} \qty( m (-5 - \cos(4\alpha)) - 2 e^{-U} \sin(2\alpha) X^3 + 6m X^4 ) \nonumber \\
-U' &= \frac{1}{4 m \sin(2\alpha) \cos(2\alpha)} \qty( m (-1 + 3 \cos(4\alpha)) - 2 e^{-U} \sin(2\alpha) X^3 + 6m X^4 ) \label{eq-2-XUodes}
\end{align}
We can note that
\( U' + \frac{X'}{X} + \frac{2\cos(2\alpha)}{\sin(2\alpha)} = 0 \)
so if we set
\( e^{-U(\alpha)} = m p X(\alpha) \sin(2\alpha) \label{eq-2-U} \)
for some (real) integration constant $p$, which can be interpreted as the curvature radius of the $AdS_2$ factor, then \eqref{eq-2-XUodes} is equivalent to solving a single ODE for $X(\alpha)$,
\( X' = \frac{X}{4 \sin(2\alpha) \cos(2\alpha)} \qty( -5 - \cos(4\alpha) + 2 \qty(3 - p \sin(2\alpha)^2) X^4) \)
The solution to this equation is
\( X = \cos(2\alpha)^{1/2} \qty(1 - p \sin(2\alpha)^2 + q \sin(2\alpha)^3)^{-1/4} \)
for some (real) integration constant $q$.
Then using Eqs.~\eqref{eq-2-Wb}, \eqref{eq-2-V}, and \eqref{eq-2-U}, we have a family of solutions to the BPS equations, labeled by real numbers $p$, $q$, and $r$,
\begin{align}
e^{2U} &= \frac{1}{p^2 m^2} \qty(1 - p \sin(2\alpha)^2 + q \sin(2\alpha)^3)^{1/2} \sin(2\alpha)^{-2} \cos(2\alpha)^{-1} \nonumber \\
e^{2W} &= \frac{1}{r^2 (3 - p)^2 m^2} \qty(1 - p \sin(2\alpha)^2 + q\sin(2\alpha)^3)^{1/2} \sin(2\alpha)^{-2} \cos(2\alpha) \nonumber \\
e^{2V} &= \frac{4}{m^2} \qty(1 - p \sin(2\alpha)^2 + q\sin(2\alpha)^3)^{-3/2} \sin(2\alpha)^{-2} \cos(2\alpha)^3 \nonumber \\
b &= \frac{1-p+q\sin(2\alpha)^3}{p^2 m^2} \sin(2\alpha)^{-1} \cos(2\alpha)^{-2} \nonumber \\
X &=  \qty( 1 - p \sin(2\alpha)^2 + q\sin(2\alpha)^3)^{-1/4} \cos(2\alpha)^{1/2}    \label{eq-2-sols}
\end{align}
In the next section we analyze how the regularity of the solutions depends on the integration constants.

\subsection{Defect solution}
\label{sec-2-2}

 The positivity of the metric factors in (\ref{eq-2-sols}) implies that the maximal  range for the coordinate $\alpha$ is  the interval $I_\alpha$ is $\alpha \in [0, \pi/4]$.
Matching the metric to that of $AdS_6$ asymptotically at the conformal boundary $\alpha \to 0$ requires equating the $e^{2U}$ and $e^{2W}$ factors, which implies
\( r^2 (3-p)^2 = p^2 \label{eq-2-rcond} \)
This should be viewed as a condition fixing $r$ in terms of $p$.
Near the conformal boundary, the $AdS_6$ radius is $\ell = m^{-1}$.
We can also observe that $X\to1$, which is the appropriate value for the global $AdS_6$ vacuum. 

The solutions \eqref{eq-2-sols} with the condition \eqref{eq-2-rcond} give a family of half-BPS solutions with $AdS_6$ asymptotics, labeled by two constants $p$ and $q$.
The $q=0$ solutions coincide with those given in \cite{Dibitetto:2018gtk}.
Incidentally, the $q=0$, $p = 1$ case describes global $AdS_6$:
\begin{align}
e^{2U} &= \frac{1}{m^2} \sin(2\alpha)^{-2}\, , & e^{2W} &= \frac{1}{m^2} \sin(2\alpha)^{-2} \cos(2\alpha)^2\, , \nonumber \\
e^{2V} &= \frac{4}{m^2} \sin(2\alpha)^{-2}\, , & b &= 0\, , \qquad\qquad  X = 1\, . \label{eq-2-ads6} 
\end{align}
Under the coordinate transformation
\( \cosh \rho = \sin(2\alpha)^{-1} \)
the metric becomes
\( \dd{s^2} = \frac{1}{m^2} \qty\Big[ \cosh^2 \rho  \dd{s^2_{AdS_2}} + \sinh^2 \rho \dd{s^2_{S^3}} + \dd{\rho^2} ] \)

\begin{figure}[h]
\centering
\includegraphics[width=.5\textwidth]{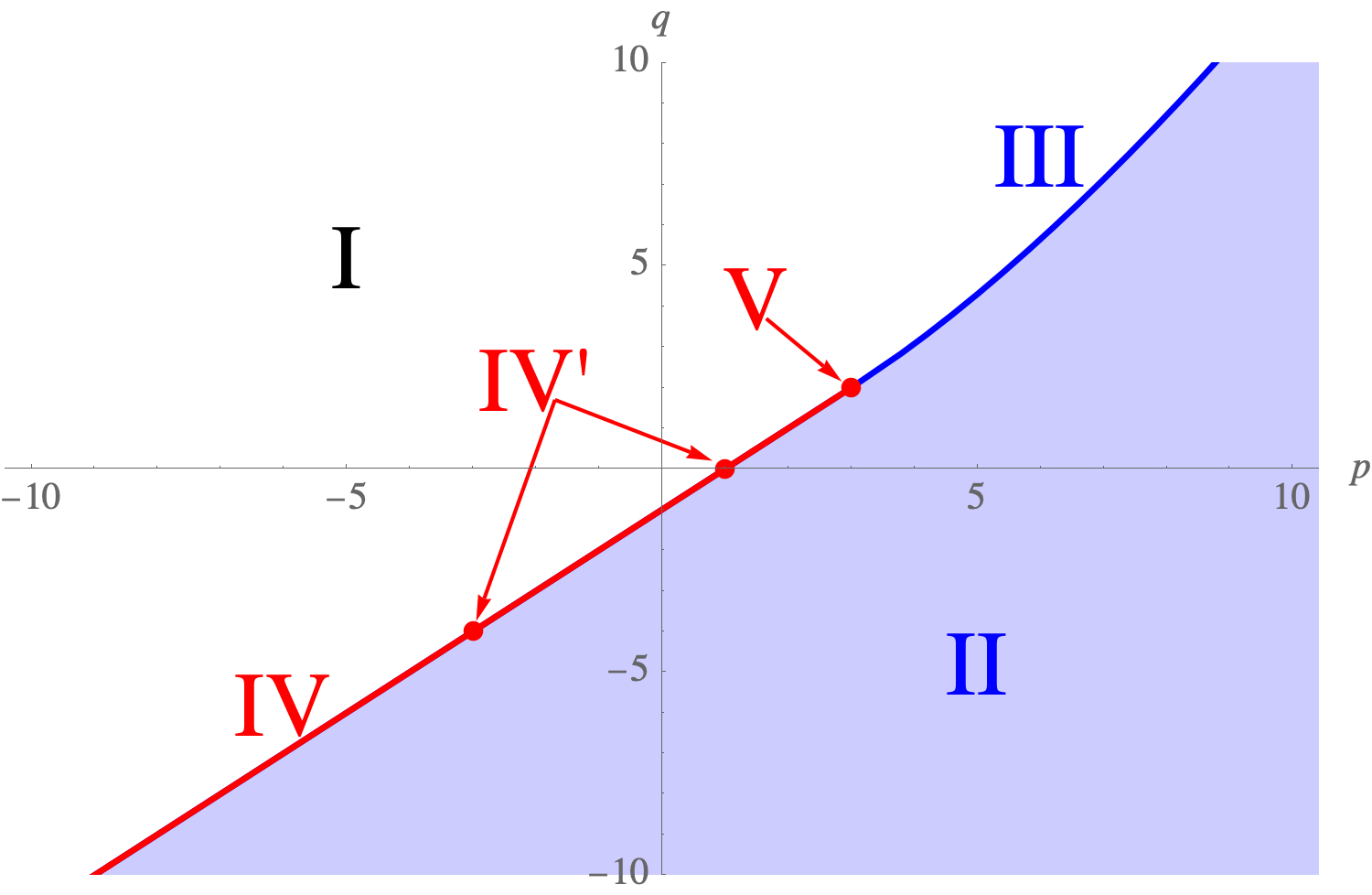}
\caption{Distinct cases shown on the $pq$-plane.}
\label{fig1}
\end{figure}

Let us now describe the behavior of these solutions at $\alpha \to \pi/4$, which corresponds to the center of the space.
The $\cos(2\alpha)$ in the metric factors vanishes here so we generically expect to have a singularity.
However, the factor
\( \Delta(\alpha) \equiv 1 - p \sin(2\alpha)^2+q \sin(2\alpha)^3 \)
may also vanish at some $0 < \alpha_0 < \pi/4$ by tuning the constants $p$ and $q$, in which case we can expect to have  a singularity located at $\alpha_0 < \pi/4$.
If we call $\beta \equiv \alpha_0 - \alpha$, then we have enough freedom to arrange for $\Delta$ to vanish as either $\order{\beta^1}$ or $\order{\beta^2}$.
We can also have $\alpha_0 = \pi/4$, in which case $\Delta$ vanishes as either $\order{\beta^2}$ or $\order{\beta^4}$ and we have to consider the $\cos(2\alpha)$ factors.
This gives us five distinct cases, which are characterized by the behavior of the metric and fields as $\beta \to 0$.
These are summarized in the table below, and the corresponding regions on the $pq$-plane are illustrated in Figure \ref{fig1}.
A ``1'' denotes approaching a constant, i.e.~$\order{\beta^0}$.

\begin{changemargin}{-1cm}{0cm} 
\[ \begin{array}{ r l | l | l l l l l | l }
&& \text{Region on }pq\text{-plane} & e^{2U} & e^{2W} & e^{2V} & B & X & \text{Ricci scalar} \\ \hline
\text{I:} &\Delta \text{ does not vanish} & \text{region I} & \beta^{-1} & \beta & \beta^3 & \beta^{-2} &\beta^{1/2} & \beta^{-5} \\
\text{II:} &\Delta \sim \beta^{\phantom{1}} \, , \quad \alpha_0 < \pi/4  & \text{region II} & \beta^{1/2} & \beta^{1/2} & \beta^{-3/2} & 1 & \beta^{-1/4} & \beta^{-1/2}\\
\text{III:} &\Delta \sim \beta^2 \, , \quad \alpha_0 < \pi/4  & q = 2(p/3)^{3/2} \text{ for } p > 3 & \beta & \beta & \beta^{-3} & 1 & \beta^{-1/2} & \beta^{-1} \quad  \\
\text{IV:} &\Delta \sim \beta^2 \, , \quad \alpha_0 = \pi/4  & q=p-1  \text{ for } p < 3 & 1 & \beta^2 & 1 & 1 & 1 & \beta^{-2} \\
\text{IV}'\text{:} &\Delta \sim \beta^2 \, , \quad \alpha_0 = \pi/4& (p, q) = (1, 0) \text{ or }(-3, -4) & 1 & \beta^2 & 1 & 1 & 1 & 1 \\
\text{V:} &\Delta \sim \beta^4 \, , \quad \alpha_0 = \pi/4  & (p, q) = (3, 2) & \beta & \beta^3 & \beta^{-3} & 1 &  \beta^{-1/2} & \beta^{-3} \\
\end{array} \]
\end{changemargin}

Case IV looks the most promising, so we will start there.
In the limit $\beta \equiv \pi/4 - \alpha \to 0$, the metric has the following leading behavior,
\( \dd{s^2} \approx \frac{32}{(6-2p)^{3/2} m^2} \qty[ \dd{\beta^2} + \frac{(6-2p)^2}{16p^2} \beta^2 \dd{s^2_{S^3}} + \frac{(6-2p)^2}{64 p^2} \dd{s^2_{AdS_2}} ] \)
We can avoid an angular deficit/excess at $\beta =0$ when $(6-2p)^2/16p^2=1$, i.e.~when $p= 1$ or $p = -3$.
These two special cases are denoted IV$'$ on the table.
The former is just global $AdS_6$, so this leaves a single nontrivial defect solution which remains finite as $\alpha \to \pi/4$, corresponding to substituting $(p, q) = (-3, -4)$ into \eqref{eq-2-sols}. 
\begin{align}
\dd{s^2} &= f_1^2 \dd{\alpha}^2  + f_2^2 \dd{s^2_{AdS_2}} + f_3^2 \dd{s^2_{S^3}}  \nonumber \\
f_1^2 &= \frac{4}{m^2} \qty( 1 + 3 \sin(2\alpha)^2 - 4 \sin(2\alpha)^3)^{-3/2} \sin(2\alpha)^{-2} \cos(2\alpha)^3 \nonumber \\
f_2^2 &= \frac{1}{9 m^2} \qty( 1 + 3 \sin(2\alpha)^2 - 4 \sin(2\alpha)^3)^{1/2} \sin(2\alpha)^{-2} \cos(2\alpha)^{-1} \nonumber \\
f_3^2 &= \frac{1}{9 m^2} \qty( 1 + 3 \sin(2\alpha)^2 - 4 \sin(2\alpha)^3)^{1/2} \sin(2\alpha)^{-2} \cos(2\alpha) \nonumber \\
b &= \frac{4}{9m^2} \qty( 1 - \sin(2\alpha)^3) \sin(2\alpha)^{-1} \cos(2\alpha)^{-2} \nonumber \\
X &=  \qty( 1 + 3 \sin(2\alpha)^2 - 4 \sin(2\alpha)^3)^{-1/4} \cos(2\alpha)^{1/2}  \label{eq-2-defect}
\end{align}
As a check, we have verified that the equations of motion \eqref{eq-1-eom} hold for this solution.   In summary, we have found a new nonsingular solution in case IV$'$, whereas all other cases I-V are singular. We will focus our analysis on the nonsingular solution
(\ref{eq-2-defect}) in the rest of the paper.

\subsection{Asymptotics}
\label{sec-2-3}

We will now calculate the asymptotic behavior of the defect solution \eqref{eq-2-defect} near the conformal boundary $\alpha \to 0$.
Recall that the $AdS_6$ radius is $\ell = m^{-1}$, which we will set to unity from here on.
Following a prescription similar to \cite{deHaro:2000vlm, Skenderis:2002wp}, we want to put the metric into the Fefferman-Graham (FG) form,
\begin{align} \dd{s^2} &= \frac{1}{z^2} \qty( \dd{z^2} + g_{ij}(x, z) \dd{x^i} \dd{x^j} ) \nonumber \\
g(x, z) &= g_0(x) + z g_1(x) + z^2 g_2(x) + \cdots \label{eq-2-FGmetric} 
\end{align}
where $i,j = 1, 2, \dotsc, 5$ run over the $AdS_2$ and $S^3$ indices, and $z \to 0$ is the conformal boundary.
This is done by taking $z = z(\alpha)$ so that the appropriate coordinate change is obtained by a solution to the ODE,
\( f_1(\alpha) \dd{\alpha} = \frac{\dd{z}}{ z} \)
Expanding in $\alpha$ and integrating term by term gives a perturbative expansion,
\begin{align}
z(\alpha) &= 3\alpha - 17 \alpha^3 + 24 \alpha^4 + \frac{722}{5} \alpha^5 - \frac{2504}{5} \alpha^6 - \frac{103009}{105} \alpha^7 + \cdots  \\
\intertext{which can be inverted,}
\alpha(z) &= \frac{1}{3} z + \frac{17}{81} z^3 - \frac{8}{81} z^4 + \frac{241}{1215} z^5 - \frac{752}{3645} z^6 - \frac{12275}{45927} z^7 + \cdots \label{eq-2-azcoord}
\end{align}
This gives the following expansions in the $z$ coordinate,
\begin{align}
f_2^2 &= \frac{1}{ z^2} \qty( \frac{1}{4} - \frac{1}{18} z^2 + \frac{1}{324} z^4 + \frac{16}{1215} z^5 + \frac{56}{2187} z^6 + \cdots ) \nonumber \\
f_3^2 &= \frac{1}{ z^2} \qty( \frac{1}{4} - \frac{1}{6} z^2 - \frac{31}{324} z^4 + \frac{32}{405} z^5 - \frac{184}{2187} z^6 + \cdots ) \nonumber \\
b &= \frac{2}{3}z^{-1} - \frac{2}{27} z + \frac{16}{81} z^3 - \frac{896}{3645} z^4 + \frac{2768}{6561} z^5 + \cdots \nonumber \\
X &= 1 - \frac{4}{9} z^2 + \frac{8}{27}z^3 - \frac{16}{81} z^4 + \frac{56}{243} z^5 - \frac{172}{729} z^6 + \frac{1072}{3645} z^7 - \frac{34304}{98415} z^8 + \cdots  \label{eq-2-expand} 
\end{align}
For the metric, we see that $g_1 = g_3 = 0$ as expected and $g_5$ will be related to the expectation value of the stress tensor.
We do not have to worry about the gravitational conformal anomaly as $d=5$ is odd, which is consistent with the fact that no terms which are logarithmic in the FG coordinate $z$ appear in the expansion. 

The conformal dimensions of the dual operators in the CFT corresponding to the scalar $\phi$ and tensor field $B$ are determined by the linearized bulk equations of motion \eqref{eq-1-eom} near the AdS boundary. 
For instance, we can plug $\phi \sim z^{\Delta_\phi}$ into the  linearized equation of motion  for the scalar in $AdS_6$ 
to obtain the relation
\( \Delta_\phi(\Delta_\phi - 5) = -6 \)
where the  $-6$ is the mass-squared of the $\phi$ field from expanding the potential $V(X)$, with $m=1$.
The mass is  within the  window  where both standard and alternative quantization is possible \cite{Klebanov:1999tb}, which implies that the scaling dimension of dual can be  either $\Delta_\phi = 2$ or  $\Delta_\phi = 3$.
However, we can argue that because the dual operators in the gravity multiplet  fall into a superconformal multiplet with the stress tensor as the top component \cite{Ferrara:1998gv}, we should have $\Delta_\phi = 3$ for the bottom scalar operator dual  to the scalar $\phi$.  It follows from the near boundary expansion (\ref{eq-2-expand}) that the defect solution has a nontrivial source as well as expectation value for the scalar operator.

Similarly, plugging $B = z^{\Delta_B-2} \dd{x^1} \wedge \dd{x^2}$ into the linearized equation of motion  for the $B$-field gives
\( (\Delta_B - 2)(\Delta_B - 3) = 2 \)
and so we have $\Delta_B = 4$ for the operator dual to 2-form potential $B$. It follows from  (\ref{eq-2-expand})  that the solution defect solution turns on a source for the operator dual to $B$.

\section{Holographic calculations}
\label{sec4}

In this section we use the formalism of holographic renormalization \cite{deHaro:2000vlm, Skenderis:2002wp} to calculate two quantities: (i) the on-shell action of the solution, which gives the expectation value of the dual defect operator, and (ii) the expectation value of the boundary stress tensor in the presence of the line defect.

\subsection{Counterterms}
\label{sec-3-1}

For a well-defined variational principle of the metric, we need to add to the bulk action given by the Lagrangian \eqref{eq-1-lagrangian} the Gibbons-Hawking boundary term,
\begin{align}
I_{\rm bulk} &= \frac{1}{16 \pi G_{\rm N}} \int_M  \mathcal{L} \nonumber \\
I_{\rm GH} &= \frac{1}{8\pi G_{\rm N}} \int_{\partial M} \dd[5]{x} \sqrt{-h} \, \Tr(h^{-1} K)
\end{align}
where $h_{ij}$ is the induced metric on the boundary and $K_{ij}$ is the extrinsic curvature.
In the FG coordinates \eqref{eq-2-FGmetric} these take the form
\begin{align}
h_{ij} &= \frac{1}{z^2} g_{ij}\,, & K_{ij} &= -\frac{ z}{2} \partial_z h_{ij}\,.
\end{align}
This action diverges due to the infinite volume of integration.
To regulate the theory, we restrict the bulk integral to the region $z \geq \epsilon$ and evaluate the boundary term at $z = \epsilon$.
Divergences in the action then appear as $1/\epsilon^k$ poles.\footnote{In  even boundary dimensions, a logarithmic term proportional to $\log \epsilon$ also appears.}
Counterterms are added on the boundary which subtract these divergent terms, leaving a renormalized action.
In all,
\( I_{\rm ren} = I_{\rm bulk} + I_{\rm GH}+ I_{\rm ct} \label{eq-3-renorm}\)
The counterterms can be expressed in terms of local quantities on the boundary.
They have been explicitly worked out in appendix \ref{appb}, which mirrors the derivation in \cite{Alday:2014bta}.\footnote{This fixes a typo in Eq.~(5.37), where the coefficient $+9/32\sqrt{2}$ should be $+7/32\sqrt{2}$ instead.}
\begin{align}
I_{\rm ct} =\, & \frac{1}{8 \pi G_{\rm N}} \int_{\partial M} \dd{x^5} \sqrt{-h} \, \bigg(  - 4 - \frac{1}{6} R[h]+ \frac{1}{8}  B^{ij}B_{ij}  \nonumber \\
& \qquad + \frac{5}{288} R[h]^2 - \frac{7}{192} R[h] B^{ij}B_{ij} + \frac{13}{512} (B^{ij}B_{ij})^2 - 4  (1-X)^2  \nonumber \\
& \qquad - \frac{1}{18} R^{ij}[h] R_{ij}[h]  - \frac{1}{6} \tensor{R}{^i_j}[h] \tensor{B}{^j_k} \tensor{B}{^k_i}  - \frac{1}{8} \tensor{B}{^i_j}\tensor{B}{^j_k}\tensor{B}{^k_\ell}\tensor{B}{^\ell_i}    \bigg) \label{eq-3-counterterms}
\end{align}
where the inverse boundary metric $h^{ij}$ is used to raise all indices and construct $R[h]$ and $R_{ij}[h]$, and $B_{ij}$ is the induced 2-form on the boundary. 
Note that this is only a subset of the most general counterterms; we have only included the terms which are nonzero for our defect solution.

Having a renormalized action allows us to obtain a finite result when computing the on-shell action of a solution.
Using the equations of motion \eqref{eq-1-eom}, we can put the on-shell ``bulk'' action into the more convenient form,
\( \eval{I_{\rm bulk}}_{\text{on-shell}} = - \frac{1}{8 \pi G_{\rm N}} \int_M X^{-2} (2 + 3 X^4) *_6 1 + \frac{1}{8 \pi G_{\rm N}} \int_{\partial M} \qty( \frac{1}{6} X^4 *_6 H \wedge B + \frac{1}{3} X^{-1} *_6 \dd{X} ) \)
The second integral over the boundary can be written more explicitly using the boundary metric $h_{ij}$ as
\( \int_{\partial M} \dd[5]{x} \sqrt{-h} \, z \qty( \frac{1}{12} X^4 B^{ij} H_{ij z} + \frac{1}{3} X^{-1} \partial_z X ) \)
The bulk integral can be performed for $\alpha \in [0, \pi/4]$ and the boundary integral, including $I_{\rm GH}$ and $I_{\rm ct}$, can be evaluated at $z = 0$.
All divergences should cancel out, by construction of the counterterms.
The on-shell action was calculated for both the global $AdS_6$ \eqref{eq-2-ads6} and defect \eqref{eq-2-defect} solutions.
\begin{align}
I_{\rm ren}(AdS_6) &= - \frac{2}{3} \cdot \frac{1}{8\pi G_{\rm N}} \text{Vol}(AdS_2) \text{Vol}(S^3) \nonumber \\
I_{\rm ren}(\text{defect}) &= \frac{2}{81} \cdot \frac{1}{8\pi G_{\rm N}} \text{Vol}(AdS_2)\text{Vol}(S^3) 
\end{align}
where $\text{Vol}(S^3) = 2\pi^2$ and $\text{Vol}(AdS_2)=-2\pi$ is the regularized  volume of $AdS_2$ \cite{Estes:2014hka,Gutperle:2016gfe}.

\subsection{Stress tensor}
\label{sec-3-2} 

Given the renormalized action, we can calculate the expectation value of the boundary stress tensor.
This contains two parts, one coming from the regularized action and one coming from the counterterms,
\( T_{ij}[h] = T_{ij}^{\rm reg}[h] + T_{ij}^{\rm ct}[h] \)
As usual, the former is given by
\( T_{ij}^{\rm reg}[h] = -\frac{2}{\sqrt{-h}} \fdv{(I_{\rm bulk} + I_{\rm GH})}{h^{ij}} = - \frac{1}{8\pi G_{\rm N}} \qty \Big(K_{ij} - h_{ij} \Tr( h^{-1} K) )  \label{eq-3-stress1}\)
The latter can be calculated by taking the variation of the counterterms in \eqref{eq-3-counterterms}, which is straightforward to compute \cite{deHaro:2000vlm,Balasubramanian:1999re}.
\( T^{\rm ct}_{ij}[h] = -\frac{2}{\sqrt{-h}} \fdv{I_{\rm ct}}{h^{ij}} \label{eq-3-stress2} \)
The expectation value of the boundary stress tensor is then related to $T_{ij}[h]$ by taking the leading term in $z$, or more concretely,
\( \ev{T_{ij}} \equiv - \frac{2}{\sqrt{-g_0}} \fdv{I_{\rm ren}}{g_0^{ij}} = \lim_{\epsilon \to 0} \qty( \epsilon^{-3} \eval{T_{ij}[h]}_{z=\epsilon} ) \)
By construction of the counterterms, this limit exists and we are left with a finite result, which we are able to write in terms of FG expansion coefficients.
Taking the following expansion of fields,
\begin{align}
z^2 h = g &= g_0 + z^2 g_2 + z^4 g_4 + z^5 g_5 + \order{z^6} \nonumber \\
B &= z^{-1} B_{-1} + z B_1 + z^2 B_2 + \order{z^3} \nonumber \\
X &= 1 + z^2 X_2 + z^3 X_3 + \order{z^4}  
\end{align}
where $B_{-1}$, $B_1$, and $B_2$ are 2-forms on the $x^1, x^2, \dotsc, x^5$ coordinates excluding $z$, the expectation value of the boundary stress tensor is
\begin{align}
\ev{T_{ij}} = \frac{1}{8\pi G_{\rm N}} \bigg[ & \frac{5}{2} {g_5}_{ij} - \frac{5}{2} {g_0}_{ij} \Tr(g_0^{-1} g_5) - \frac{1}{4} {g_0}_{ij} \Tr(g_0^{-1} B_{-1} g_0^{-1} B_2) \nonumber \\
& + \frac{1}{2} {B_{-1}}_{ik} g_0^{k\ell} {B_2}_{\ell j} + \frac{1}{2}{B_{2}}_{ik} g_0^{k\ell} {B_{-1}}_{\ell j} - 8 {g_0}_{ij} X_2 X_3 \bigg]
\end{align}
This quantity depends on the FG coefficients left undetermined by the equations of motion, namely $g_5$, $B_2$, and $X_3$, as expected
Taking the trace with the conformal boundary metric $g_0$ gives,
\( \ev{T^i_i} = \frac{1}{8 \pi G_{\rm N}} \qty[ -10 \Tr(g_0^{-1} g_5) - \frac{1}{4} \Tr(g_0^{-1} B_{-1} g_0^{-1} B_2) - 40 X_2 X_3 ] \)
This result is accompanied by a Ward identity encoding the spontaneous breaking of scale invariance,
\( \frac{5}{2} \Tr(g_0^{-1} g_5) + 12 X_2 X_3 + \frac{1}{4} \Tr(g_0^{-1} B_{-1} g_0^{-1} B_2) - \frac{1}{4} X_3 \Tr( g_0^{-1} B_{-1} g_0^{-1} B_{-1}) = 0 \)
which comes from the bulk Einstein equation \eqref{eq-1-eom}, expanded in FG coordinates to order $\order{z^3}$. 
Explicitly evaluating these two expectation values for our defect solution, using the expansion coefficients in \eqref{eq-2-expand}, yields
\begin{align}
\ev{T_{ij}} &= \mqty( - \frac{88}{243} g_{AdS_2} & 0 \\ 0 & -\frac{16}{81} g_{S^3} ) \,, & \ev{T^i_i} &= -\frac{1280}{243}\,,
\end{align}
where $g_{AdS_2}$ and $g_{S^3}$ are unit radius.

\section{Discussion}
\label{sec5}

In this paper we found  a nonsingular solution of  $F(4)$ gauged supergravity, which is of the form $AdS_2\times S^3$ warped over an interval. It preserves eight  of the sixteen supersymmetries and represents a holographic dual of a half-BPS superconformal line defect.
This solution is uniquely determined by the symmetries of the ansatz and the fact that it is half-BPS. Solutions of  $F(4)$ gauged supergravity can be consistently lifted to $AdS_6$ solutions of massive IIA   \cite{Cvetic:1999un} or type IIB solutions \cite{Malek:2018zcz,Hong:2018amk}. Consequently, the solution found  in this paper lifts to a holographic line defect for the ten-dimensional theories. The ten-dimensional warped $AdS_6$ solutions have a holographic field theory dual such as $USp(N)$ gauge theories for massive type IIA and long quiver theories coming from $(p,q)$ five-brane webs for type IIB.

The lifted  solution should correspond to a heavy line defect in these ten-dimensional theories and is universal in the sense that it exists in all of then dimensional $AdS_6$ solutions. However, unlike  the holographic Wilson line solutions for $N=4$ SYM found in \cite{DHoker:2007mci},  we do not know which representation the line defect corresponds to and we do not have families of solutions corresponding to different representations in a given $AdS_6$ vacuum. 
One way to obtain such solutions is to start in the ten-dimensional theory, but since even the $AdS_6$ vacuum  has the form of a warped product this is considerably harder than in the $AdS_5\times S^5$ case. 
The form of the lifted solution may give hints on how a more general ansatz should look like. Furthermore, generalizing the solution found in this paper to theories which include additional vector multiplets may be useful, since a consistent truncation in some cases  was found recently \cite{Malek:2019ucd}. We leave these interesting questions for future work.

\section*{Acknowledgements}
The work of M.G.~is supported in part by the National Science Foundation under grant PHY-19-14412. 
K.C.~is grateful to the Bhaumik Institute for Theoretical Physics for support.

\newpage

\appendix

\section{Conventions}
\label{appa}

The six-dimensional Hodge dual is given by
\( *_6 (\dd{x^{\mu_1}} \wedge \cdots \wedge \dd{x^{\mu_r}} ) = \frac{\sqrt{-G}}{(6-r)!} \tensor{\epsilon}{_{\nu_1 \dotsc \nu_{D-r}}^{\mu_1 \dotsc \mu_r}} \dd{x^{\nu_1}} \wedge \cdots \dd{x^{\nu_{D-r}}} \)
where $\epsilon_{123456} = 1$.
More concretely, we use the coordinates
\[ \begin{array}{r | c  c | c c c | c}
 & \multicolumn{2}{|c| }{AdS_2} & \multicolumn{3}{|c|}{S^3} & I_\alpha \\ \hline
\mu = & 1 & 2 & 3 & 4 & 5 & 6 \\ \hline
x^\mu = & t & r & \psi & \theta & \phi & \alpha \text{ or } z
\end{array} \]
\begin{align}
\dd{s^2_{AdS_2}} &= -(1 + r^2) \dd{t^2} + (1 + r^2)^{-1} \dd{r^2} \nonumber \\
\dd{s^2_{S^3}} &= \dd{\psi^2} + \sin^2\psi \dd{\theta^2} + \sin^2\psi \sin^2\theta \dd{\phi^2}
\end{align}
The norm of a $p$-form is defined as
\( \norm{F}_g^2 = \frac{1}{p!} F^{\mu_1 \dotsc \mu_p} F_{\mu_1 \dotsc \mu_p} \)
where all indices are raised using the specified metric $g$.
For the Riemann curvature tensor, we use the sign convention
\begin{align}
\tensor{R}{^\rho_{\sigma \mu\nu}} &= \partial_\mu \Gamma^\rho_{\nu\sigma} + \Gamma^\rho_{\mu \lambda} \Gamma^\lambda_{\nu \sigma} - (\mu \leftrightarrow \nu) \nonumber \\
R_{\mu\nu} &= \tensor{R}{^\rho_{\mu\rho\nu}}
\end{align}

\section{Counterterms}
\label{appb}

Here we briefly outline the calculation for obtaining the counterterms in \eqref{eq-3-counterterms}.
We will follow the steps in \cite{Alday:2014bta}, making some simplifications suited for our purposes. 

Using the Einstein equation \eqref{eq-1-eom} we can write the on-shell bulk action as
\( \eval{I_{\rm bulk}}_{\text{on-shell}} = \frac{1}{16 \pi G_{\rm N}} \int_M \qty[ \frac{1}{2} V(X) *_6 1 - \frac{1}{2} X^4 *_6 H \wedge H - \frac{1}{2} X^{-2} *_6 B\wedge B -\frac{1}{3} B \wedge B \wedge B ] \label{eq-B-action1} \)
where we have set $m=1$ and ignored terms involving $A^i$ and $A^0$.
The on-shell action also includes the Gibbons-Hawking term,
\( I_{\rm GH} = \frac{1}{16\pi G_{\rm N}} \int_{\partial M} \dd[5]{x} \qty(-2 z \partial_z \sqrt{-h}\, ) \label{eq-B-action2} \)
We assume the following expansions of the fields,
\begin{align}
g &= g_0 + z^2 g_2 + z^4 g_4 + \order{z^5} \nonumber \\
B &= z^{-1} B_{-1} + \dd{z} \wedge A_0 + z B_1 + \order{z^2} \nonumber \\
H &= - z^{-2} \dd{z} \wedge B_{-1} + z^{-1} \dd{B_{-1}} - \dd{z} \wedge \dd{A_0} + \dd{z} \wedge B_1 + \order{z} \nonumber \\
X &= 1 + z^2 X_2 + \order{z^3}  
\end{align}
where $B_{-1}$ and $B_1$ are 2-forms on the $x^1, x^2, \dotsc, x^5$ coordinates excluding $z$, and $A_0$ is a 1-form on the same coordinates.
The general strategy is to plug these expansions into the on-shell action, integrate the bulk terms over $z \geq \epsilon$, and evaluate the boundary terms at $z = \epsilon$.
We will have order $\order{\epsilon^{-5}}$, $\order{\epsilon^{-3}}$, and $\order{\epsilon^{-1}}$ divergences, which are worked out order-by-order and then canceled out by appropriate counterterms.
It is important to remember that the counterterm added to cancel the $\order{\epsilon^{-5}}$ divergence will also contribute to the $\order{\epsilon^{-3}}$ divergence, and so  forth. \\

Along the way, we will need to use the equations of motion \eqref{eq-1-eom} expanded in the FG coordinates \eqref{eq-2-FGmetric}.
This requires the expansion of the six-dimensional Ricci tensor in these coordinates, 
\begin{align}
R_{zz} &= \frac{1}{4} \Tr( g^{-1} g' g^{-1} g' ) - \frac{1}{2} \Tr(g^{-1} g'') + z^{-1} \frac{1}{2} \Tr(g^{-1} g') - 5 z^{-2} \nonumber \\
R_{iz} &= \frac{1}{2} g^{jk} \nabla_k g_{ij}' - \frac{1}{2} g^{jk} \nabla_i g_{jk}' \nonumber \\
R_{ij} &= \frac{1}{2} g_{ik}' g^{k\ell} g_{\ell j}' - \frac{1}{4} g_{ij}' \Tr(g^{-1} g') - \frac{1}{2} g_{ij}'' + R_{ij}[g] + z^{-1} \qty( 2 g_{ij}' + \frac{1}{2} g_{ij} \Tr(g^{-1} g') ) - 5 z^{-2} g_{ij} 
\end{align}
where $R_{ij}[g]$ and $\nabla_i$ are constructed using the five-dimensional metric $g = g_0 + g_1 z + g_2 z^2 + \cdots$, and $'$ denotes the derivative with respect to $z$.
For instance, the order $\order{z^{0}}$ Einstein equation implies that
\begin{align}
{g_2}_{ij} &= - \frac{1}{3} \qty( R_{ij}[g_0]  - \frac{1}{8} {g_0}_{ij} R[g_0] ) - \frac{3}{16} {g_0}_{ij} \norm{B_{-1}}^2_{g_0} - \frac{1}{2} {B_{-1}}_{ik} {g_0}^{k\ell} {B_{-1}}_{\ell j} \nonumber \\
\Tr( g_0^{-1} g_2) &= - \frac{1}{8} R[g_0] + \frac{1}{16} \norm{B_{-1}}_{g_0}^2 \label{eq-B-g2}
\end{align}
Another useful expansion is the determinant,
\( \sqrt{-g} = \sqrt{-g_0}\, \qty[ 1 + \frac{1}{2} z^2 \Tr(g_0^{-1} g_2) + \frac{1}{2} z^4 \qty( \Tr(g_0^{-1} g_4) - \frac{1}{2} \Tr(g_0^{-1} g_2 g_0^{-1} g_2) + \frac{1}{4} \Tr^2(g_0^{-1} g_2) ) + \cdots ] \) 

For each order in $\epsilon$, we will give the contributing divergence from each term in the action (\ref{eq-B-action1}, \ref{eq-B-action2}), omitting an implicit $\sqrt{-g_0}/16 \pi G_{\rm N}$ factor. \\

\textbf{Order $\order{\epsilon^{-5}}$:}
\begin{align*}
\frac{1}{2} V(X) *_6 1 &:& -2  \\
-2 z \partial_z \sqrt{-h} &:& 10 
\end{align*}
Adding these two contributions and restoring the $\sqrt{-g_0}/16 \pi G_{\rm N}$ factor, the $\order{\epsilon^{-5}}$ divergence of the on-shell action is
\(  I_5 = \frac{\epsilon^{-5}}{16 \pi G_{\rm N}} \int_{z = \epsilon}\dd[5]{x} \sqrt{-g_0} \, 8 \)
A suitable counterterm which cancels this at leading order is
\( I_{\rm ct, 5} = \frac{1}{16 \pi G_{\rm N}} \int_{\partial M} \dd[5]{x} \sqrt{-h} \, (-8) \) \\

\textbf{Order $\order{\epsilon^{-3}}$:}
\begin{align*}
\frac{1}{2} V(X) *_6 1 &:& -\frac{5}{3} \Tr(g_0^{-1} g_2)  \\
-\frac{1}{2} X^{4} *_6 H \wedge H &:& -\frac{1}{6} \norm{B_{-1}}^2_{g_0} \\
-\frac{1}{2} X^{-2} *_6 B \wedge B &:& -\frac{1}{6} \norm{B_{-1}}^2_{g_0} \\
-2 z \partial_z \sqrt{-h} &:&  3 \Tr(g_0^{-1} g_2)    \\
-8 \sqrt{-h} &:& -4 \Tr(g_0^{-1} g_2) 
\end{align*}
\( \implies I_3 = \frac{\epsilon^{-3}}{16 \pi G_{\rm N}} \int_{z = \epsilon}\dd[5]{x} \sqrt{-g_0} \, \qty( \frac{1}{3} R[g_0]  - \frac{1}{2} \norm{B_{-1}}^2_{g_0} ) \)
where we used  \eqref{eq-B-g2}.
Thus,
\( I_{\rm ct, 3} = \frac{1}{16 \pi G_{\rm N}} \int_{\partial M} \dd[5]{x} \sqrt{-h} \, \qty( - \frac{1}{3} R[h] + \frac{1}{2} \norm{B}^2_h ) \)

In order to write down the $\order{\epsilon^{-1}}$ divergences, we need the FG expansion of $R[h] = z^2 R[g]$.
A particularly convenient expansion is obtained from the order $\order{z^2}$ Einstein equation, which implies that
\begin{align}
R[g] =\, & R[g_0] + z^2 \bigg( - 8 \Tr (g_0^{-1} g_4) + 5 \Tr (g_0^{-1} g_2 g_0^{-1} g_2) + \Tr^2 (g_0^{-1} g_2)- 20 {X_2}^2 - X_2 \norm{B_{-1}}^2_{g_0}   \nonumber \\
& \qquad + \frac{1}{2} \Tr (g_0^{-1} B_{-1} g_0^{-1} B_{-1} g_0^{-1} g_2) - \frac{1}{2} \Tr (g_0^{-1} B_{-1} g_0^{-1} B_1) + \frac{1}{2} \norm{A_0}^2_{g_0}  \bigg) + \order{z^4} \nonumber \\
\Tr(g_0^{-1} g_4) =\, & \frac{1}{4} \Tr( g_0^{-1} g_2 g_0^{-1} g_2)  - \frac{5}{2} {X_2}^2 - \frac{3}{8} X_2 \norm{B_{-1}}^2_{g_0} - \frac{1}{8} \Tr(g_0^{-1} B_{-1} g_0^{-1} B_1) \nonumber \\
&\qquad  + \frac{1}{16} \Tr(g_0^{-1} B_{-1} g_0^{-1} \dd{A_0}) - \frac{3}{16} \norm{A_0}^2_{g_0} + \frac{1}{16} \norm{\dd{B_{-1}}}^2_{g_0} \label{eq-B-g4}
\end{align}
We will also further assume $A_0 = 0$ and $\dd{B_{-1}} = 0$, which is not true in general but is true for our solution and vastly simplifies calculations. \\

\begin{changemargin}{-1cm}{0cm} 
\hspace{1.5cm} \textbf{Order $\order{\epsilon^{-1}}$:}
\begin{align*}
\frac{1}{2} V(X) *_6 1 &:& - 5 \Tr(g_0^{-1} g_4) + \frac{5}{2} \Tr(g_0^{-1} g_2 g_0^{-1} g_2) - \frac{5}{4} \Tr^2(g_0^{-1} g_2) - 12 {X_2}^2   \\
-\frac{1}{2} X^{4} *_6 H \wedge H &:& -\frac{1}{4} \qty( \Tr(g_0^{-1} g_2) + 8X_2) \norm{B_{-1}}^2_{g_0} - \frac{1}{2} \Tr (g_0^{-1} B_{-1} g_0^{-1} B_{-1} g_0^{-1} g_2) - \frac{1}{2} \Tr(g_0^{-1} B_{-1} g_0^{-1} B_1)  \\
-\frac{1}{2} X^{-2} *_6 B \wedge B &:&  -\frac{1}{4} \qty( \Tr(g_0^{-1} g_2) - 4 X_2) \norm{B_{-1}}^2_{g_0} - \frac{1}{2} \Tr (g_0^{-1} B_{-1} g_0^{-1} B_{-1} g_0^{-1} g_2)  + \frac{1}{2} \Tr(g_0^{-1} B_{-1} g_0^{-1} B_1 )  \\
-2 z \partial_z \sqrt{-h} &:& \Tr(g_0^{-1} g_4) - \frac{1}{2} \Tr(g_0^{-1} g_2 g_0^{-1} g_2) + \frac{1}{4} \Tr^2(g_0^{-1} g_2) \\
-8 \sqrt{-h} &:&  -4 \Tr(g_0^{-1} g_4) + 2 \Tr(g_0^{-1} g_2 g_0^{-1} g_2) - \Tr^2(g_0^{-1} g_2) \\
-\frac{1}{3} R[h] \sqrt{-h} &:&  \frac{8}{3} \Tr(g_0^{-1} g_4) - \frac{5}{3} \Tr( g_0^{-1} g_2 g_0^{-1} g_2) + \Tr^2(g_0^{-1} g_2)  + \frac{20}{3} {X_2}^2 - \frac{1}{12} \norm{B_{-1}}^2_{g_0} \Tr(g_0^{-1} g_2)  \qquad  \\
&&  + \frac{1}{3} X_2 \norm{B_{-1}}^2_{g_0} - \frac{1}{6} \Tr( g_0^{-1} B_{-1} g_0^{-1} B_{-1} g_0^{-1} g_2) + \frac{1}{6} \Tr(g_0^{-1} B_{-1} g_0^{-1} B_1)  \\
\frac{1}{2} \norm{B}^2_h \sqrt{-h} &:&  \frac{1}{4} \norm{B_{-1}}^2_{g_0} \Tr( g_0^{-1} g_2)+\frac{1}{2} \Tr( g_0^{-1} B_{-1} g_0^{-1} B_{-1} g_0^{-1} g_2) - \frac{1}{2} \Tr( g_0^{-1} B_{-1} g_0^{-1} B_1)  
\end{align*}
\begin{align}
\implies I_1 =\, & \frac{\epsilon^{-1}}{16 \pi G_{\rm N}} \int_{z = \epsilon}\dd[5]{x} \sqrt{-g_0} \, \bigg( - \frac{5}{144} R[g_0]^2 + \frac{7}{48} \norm{B_{-1}}^2_{g_0} R[g_0] - \frac{13}{64} \norm{B_{-1}}^4_{g_0} + 8 {X_2}^2 \nonumber \\
& \qquad + \frac{1}{9} \Tr( g_0^{-1} \text{Ric}[g_0] g_0^{-1} \text{Ric}[g_0] ) + \frac{1}{3} \Tr( g_0^{-1} \text{Ric}[g_0]  g_0^{-1}  B_{-1}  g_0^{-1}  B_{-1} ) + \frac{1}{4} \Tr[( g_0^{-1}  B_{-1})^4] \nonumber \\
& \qquad - \frac{1}{3} \norm{B_{-1}}^2_{g_0} \Tr(g_0^{-1} g_2) + \frac{4}{3} X_2 \norm{B_{-1}}^2_{g_0} - \frac{2}{3} \Tr( g_0^{-1}  B_{-1} g_0^{-1}  B_{-1} g_0^{-1}  g_2) + \frac{1}{3} \Tr( g_0^{-1}  B_{-1} g_0^{-1}  B_1) \bigg)
\end{align}
\end{changemargin}
where we used \eqref{eq-B-g2} and \eqref{eq-B-g4}.
The terms on the last line cancel out using the order $\order{z^{-1}}$ $B$-field equation of motion, which implies
\( {B_1}_{ij} = 2 X_2 {B_{-1}}_{ij} - \frac{1}{2} \Tr(g_0^{-1} g_2) {B_{-1}}_{ij} + {g_2}_{ik} g_0^{k\ell} {B_{-1}}_{\ell j} + {B_{-1}}_{ik} g_0^{k\ell} {g_2}_{\ell j} \)
Thus, a suitable choice of counterterms is
\begin{align}
I_{\rm ct, 1} =\, & \frac{1}{16 \pi G_{\rm N}} \int_{\partial M} \dd[5]{x} \sqrt{-h} \, \bigg( \frac{5}{144} R[h]^2 - \frac{7}{48} \norm{B}^2_h R[h] + \frac{13}{64} \norm{B}^4_h - 8 (1-X)^2 \nonumber \\
& \qquad - \frac{1}{9} \Tr( h^{-1} \text{Ric}[h] h^{-1} \text{Ric}[h] ) - \frac{1}{3} \Tr( h^{-1} \text{Ric}[h]  h^{-1}  B  h^{-1}  B ) - \frac{1}{4} \Tr[ ( h^{-1}  B)^4]  \bigg) 
\end{align}
This fixes a typo in Eq.~(5.37) of \cite{Alday:2014bta}, where the coefficient $+9/32\sqrt{2}$ should be $+7/32\sqrt{2}$ instead.

\newpage
\providecommand{\href}[2]{#2}\begingroup\raggedright\endgroup

\end{document}